\documentclass[%
amsmath,amssymb,
reprint,
]{revtex4-1}

\usepackage{graphicx}
\usepackage{dcolumn}
\usepackage{bm}

\usepackage{amsmath}
\usepackage{color}
\usepackage[normalem]{ulem}
\usepackage{balance}
\usepackage[T1]{fontenc}

\makeatletter
\renewcommand*{\@fnsymbol}[1]{\ifcase#1\or*\else\@arabic{#1}\fi}
\makeatother

\begin{document}
\title{Mechanisms of chiral plasmonics - scattering, absorption and photoluminescence}

\author{Yuqing Cheng}\affiliation{School of Mathematics and Physics, University of Science and Technology Beijing, Beijing 100083, China}%
\author{Mengtao Sun}\thanks{Corresponding author: mengtaosun@ustb.edu.cn}
\affiliation{School of Mathematics and Physics, University of Science and Technology Beijing, Beijing 100083, China}%
\begin{abstract}
\textbf{ABSTRACT: } Chirality is a concept that one object is not superimposable on its mirror image by translation and rotation. In particular, chiral plasmonics have been widely investigated due to their excellent optical chiral properties, and have led to numerous applications such as optical polarizing element etc. In this study, we develop a model based on the concept of the interaction between harmonic oscillators to investigate and explain the optical chiral mechanisms of strongly coupled metal nanoparticles (MNPs). The chirality of the scattering, absorption, and photoluminescence spectra are carefully discussed in detail. The results show that the chirality of the system originates not only from the orientations of the MNPs, but also from the different eigen parameters between them. Specifically, the derived three factors contribute to the chirality: the symmetry, the coupling strength, and the coherent superposition of the emitted electric field. This work provides a deeper understanding on the chiral plasmonics and may guide relevant applications in theory.
\end{abstract}
\maketitle

\section{\label{sec:Introduction}Introduction}
An object has chirality if it is not superimposable on its mirror image by translation and rotation. Usually, a chiral object is lack of symmetry such as mirror planes, inversion centers or improper rotational axes \cite{chirality,Chirality2}.
In particular, chirality in the field of plasmonics has attracted numerous attentions due to its promising applications, such as light controlling \cite{appLC1,appLC2,appLC3,appLC4},  polarization-sensitive optical devices \cite{appDevice1,appDevice2,appDevice3,appDevice4,appDevice5,appDevice6}, molecular sensing \cite{appMS1,appMS2,appMS3,appMS4}, chiral catalysis \cite{appCC1}.

Recently, Li Z. $et~al.$ \cite{Science} apply the chiral quadrupole field introduced by permanent magnets to dispersed magnetic nanoparticles, which results in long-range chiral superstructures. By tuning the magnetic field, they are able to control over the superstructures' handedness and chiroptical properties. Moreover, the chirality could be transferred to organic molecules and inorganic compounds.
Singh G. $et~al.$ \cite{Science2014} use the self-assembly technology to synthesize different types of superstructures composed of magnetic nanocubes, such as belts, as well as single, double, and triple helices. They reveal a novel mechanism of symmetry breaking and chirality amplification, i.e., the neighboring helices tend to arrange in the same handedness thus maximizing packing.
Jeong K. $et~al.$ \cite{ACSnano2020} employ magnetoplasmonic nanoparticles to guide plasmonic Ag nanoparticles onto a helical magnetic flux. They are successful in tuning the chirality of the structures in real time, thus controlling the polarization state of light.

All the above works are excellent in controlling light by introducing the chirality of the structures in different ways. However, a deeper understanding in the physical mechanisms of the chirality of these complex structures is in demand. In this study, we develop a model to investigate and explain the optical chiral mechanisms of strongly coupled metal nanoparticles (MNPs). This model is based on the concept of the interaction between harmonic oscillators. The chirality of the scattering, absorption, and photoluminescence (PL) spectra of the coupled MNPs are investigated in detail.
We only consider two MNPs so that the mechanisms can be revealed as simple as possible. Further investigation on the chirality of multiple MNPs will be proceeded in the future.

\section{\label{sec:Model}Model}
\begin{figure}[tb]
\includegraphics[width=0.48\textwidth]{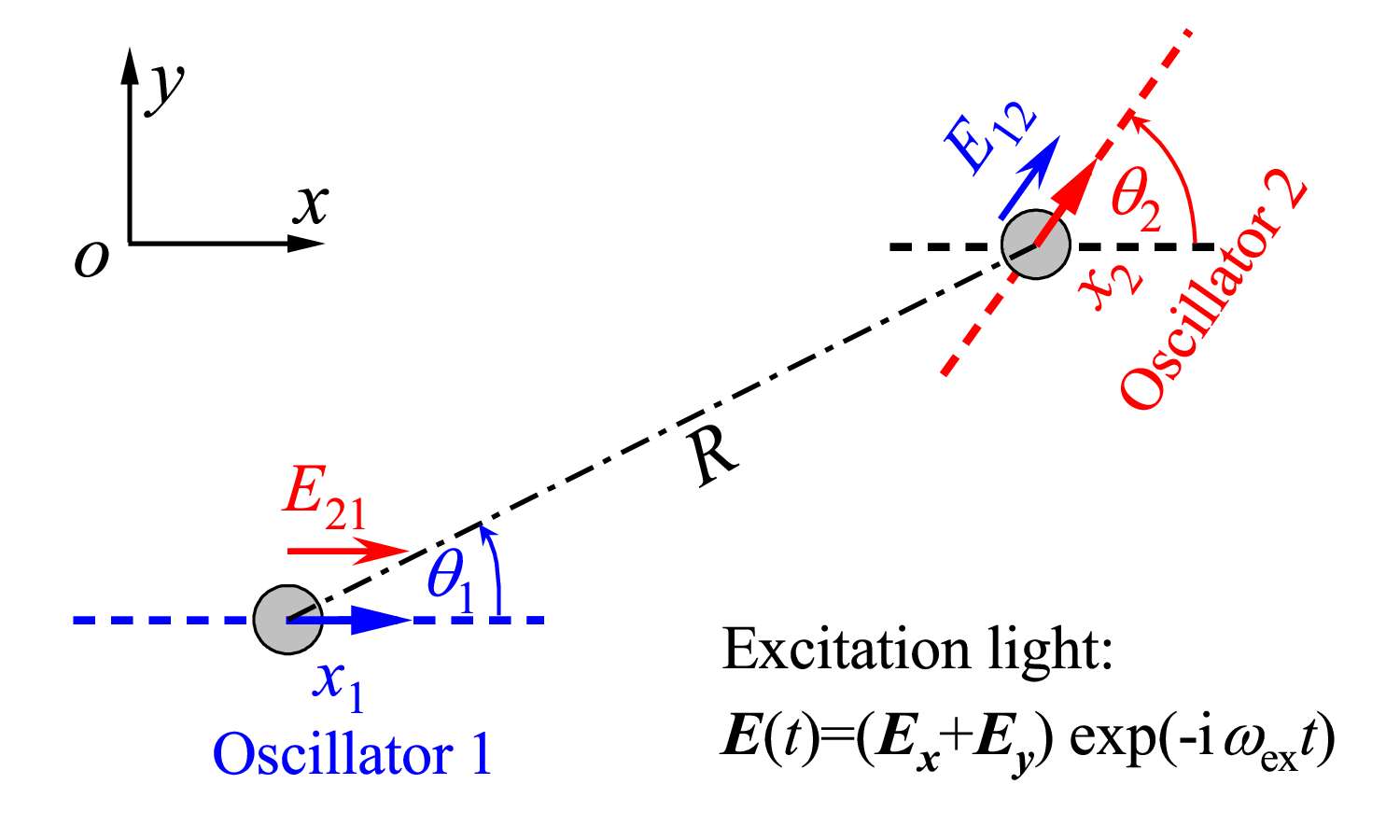}
\caption{\label{fig:Scheme} Schematic of the coupled MNPs (oscillators). The distance between the two oscillators is $R$. The angle between the link line and the $x$-axis is $\theta_1$. The mode of oscillator 1 is paralleled to the $x$-axis, and the angle between the mode of oscillator 2 and the $x$-axis is $\theta_2$. $E_{12}$ ($E_{21}$) is the electric field introduced by oscillator 1 (2) felt by oscillator 2 (1). The excitation light propagates along the $z$-axis with the angular frequency of $\omega_{ex}$, and $E_x$ and $E_y$ are the components of the external electric field in the $x$ and $y$ directions, respectively.
}
\end{figure}

As we have presented in our previous work \cite{couplex,coupley}, we treat each individual MNP as an oscillator with their own eigen angular frequency $\omega_j$ and damping $\beta_j$. The oscillator is composed of numerous free electrons which are oscillating collectively when excited by external light. Here, the oscillating modes of the two coupled MNPs are in arbitrary directions. The schematic is shown in Fig. \ref{fig:Scheme}. We use $x_j$ ($j=1,2$) to present the displacements of the $j$th oscillator, the directions of which are along their mode directions, respectively. In general, $\dot{x}_j$ and $\ddot{x}_j$ represent the corresponding velocity and acceleration. Therefore, the equations of the two oscillators are written as:
\begin{equation}
\begin{aligned}
&\ddot{x}_1+\beta_1 \dot{x}_1 +\omega^2_1 x_1 + \frac{e E_{21}}{m_e}=C_1 \mathrm{exp}(-\mathrm{i} \omega_{ex}t), \\
&\ddot{x}_2+\beta_2 \dot{x}_2 +\omega^2_2 x_2 + \frac{e E_{12}}{m_e} =C_2 \mathrm{exp}(-\mathrm{i} \omega_{ex}t).
\end{aligned}
\label{eq:couple0}
\end{equation}
Here, $e$ is the elementary charge, $m_e$ is the mass of electron, $C_1=-\frac{e}{m_e} E_x$, $C_2=-\frac{e}{m_e} (E_x \cos \theta_2 +E_y \sin \theta_2)$, and $E_x$ and $E_y$ are the electric field of the incident light in $x$ and $y$ directions, respectively.
The electric field introduced by the two oscillators can be evaluated by \cite{Griffiths}:
\begin{equation}
\begin{aligned}
&E_{12}\cong -\frac{eN_1}{4 \pi \varepsilon_0 R^2}
\left( \Theta_1 \frac{x_1}{R} +\Theta_1 \frac{\dot{x}_1}{c} + \Theta_2 \frac{R\ddot{x}_1}{c^2} \right), \\
&E_{21}\cong -\frac{eN_2}{4 \pi \varepsilon_0 R^2}
\left( \Theta_1 \frac{x_2}{R} +\Theta_1 \frac{\dot{x}_2}{c} + \Theta_2 \frac{R\ddot{x}_2}{c^2} \right).
\end{aligned}
\label{eq:E12}
\end{equation}
Here, $N_1$ and $N_2$ are the free electron numbers of oscillator 1 and 2, respectively; $\varepsilon_0$ is the permittivity of free space, $R$ is the distance between the two MNPs, and $c$ is the speed of light in vacuum; the parameters $\Theta_1$ and $\Theta_2$ are functions of $(\theta_1,\theta_2)$, and they are written as:
\begin{equation}
\begin{aligned}
& \Theta_1(\theta_1,\theta_2)=3\cos \theta_1 \cos (\theta_2-\theta_1) -\cos \theta_2,\\
& \Theta_2(\theta_1,\theta_2)=\sin \theta_1 \sin (\theta_2-\theta_1).
\end{aligned}
\label{eq:Th12}
\end{equation}
Therefore, Eq. \ref{eq:couple0} can be written as:
\begin{equation}
\begin{aligned}
&\ddot{x}_1+\beta_1 \dot{x}_1 +\omega^2_1 x_1
+ \eta_2 \ddot{x}_2 + \gamma_2 \dot{x}_2 + g_2^2 x_2  =C_1 \mathrm{exp}(-\mathrm{i} \omega_{ex}t), \\
&\ddot{x}_2+\beta_2 \dot{x}_2 +\omega^2_2 x_2
+ \eta_1 \ddot{x}_1 + \gamma_1 \dot{x}_1 + g_1^2 x_1 =C_2 \mathrm{exp}(-\mathrm{i} \omega_{ex}t).
\end{aligned}
\label{eq:couple1}
\end{equation}
Here, the coupling coefficients are defined as:
\begin{equation}
\begin{aligned}
&g_j^2= \frac{- N_j e^2 \Theta_1}{4 \pi \varepsilon_0 m_e R^3},
~~~~\gamma_j= \frac{- N_j e^2 \Theta_1}{4 \pi \varepsilon_0 m_e R^2 c},\\
&\eta_j= \frac{-N_j e^2 \Theta_2}{4 \pi \varepsilon_0 m_e R c^2},
~~~~j=1,2.
\end{aligned}
\label{eq:CS} 
\end{equation}
The form of Eq. \ref{eq:couple1} is the same as Eq. 6 of Ref. \cite{coupley}. The difference is that the coupling coefficients in Eq. \ref{eq:couple1} are additionally dependent on $\theta_1$ and $\theta_2$, but the ones of the latter are corresponding to the particular case of $(\theta_1,\theta_2)=(\pm \pi/2,0)$. Furthermore, Ref. \cite{couplex} is a specific case of $(\theta_1,\theta_2)=(0,0)$.
First, we derive the formulas of the scattering, absorption, and PL spectra of the system.
Next, we investigate the chirality of the strong coupled MNPs by analysing these spectra when excited by the right-handed (RCP) and left-handed circularly polarized (LCP) light, respectively.
Last, we show how are the spectra are controlled by varying the orientation of the MNPs.
For simplicity, we define $\Omega_j(\alpha)=\omega_j^2+\beta_j^2 \alpha+\alpha^2$ and $G_j(\alpha)=g_j^2+\gamma_j \alpha +\eta_j \alpha^2$ for $j=1,\ 2$, which would be used in the following derivations.

\subsection{\label{sec:Sca}Scattering and absorption}
For white light scattering spectra, $\alpha=-\mathrm{i}\omega_{ex}$. Assume $x_j(t)=A_j \mathrm{exp}(-\mathrm{i}\omega_{ex}t)$ for $j=1,\ 2$, and substitute them into Eq. \ref{eq:couple1}, we obtain the equation:
\begin{equation}
\begin{aligned}
&
\begin{pmatrix}
\Omega_1 & G_2 \\
G_1 & \Omega_2
\end{pmatrix}
\left(
\begin{array}{c}
A_1 \\
A_2 \\
\end{array}\right)
=\left(
\begin{array}{c}
C_1 \\
C_2 \\
\end{array}\right),
\\ & \mathrm{with}~ \alpha=-\mathrm{i}\omega_{ex}.
\end{aligned}
\label{eq:SCAm}
\end{equation}
The solutions are:
\begin{equation}
\begin{aligned}
A_1(\omega_{ex})=\frac{\Omega_2 C_1-G_2 C_2}{\Omega_1 \Omega_2 - G_1 G_2},\\
A_2(\omega_{ex})=\frac{\Omega_1 C_2-G_1 C_1}{\Omega_1 \Omega_2 - G_1 G_2}.
\end{aligned}
\label{eq:A}
\end{equation}
The scattering spectrum is usually related to the far field, therefore the electric field is approximately proportional $\ddot{x}_j \propto \omega_2$. After substituting $\omega_{ex}$ with $\omega$ we obtain the white light scattering spectrum:
\begin{equation}
\begin{aligned}
I_{sca}(\omega)=&\omega^4  \left|N_1 A_1(\omega)+N_2 A_2(\omega) \cos \theta_2 \right|^2
\\+& \omega^4 \left|N_2 A_2(\omega) \sin \theta_2 \right|^2 .
\end{aligned}
\label{eq:Isca}
\end{equation}
The absorption spectrum is introduced by the term $\beta_j \dot{x}_j$ which refers to the friction force on the electrons \cite{Fano}. Hence, the absorption spectrum can be written as:
\begin{equation}
\begin{aligned}
I_{abs}(\omega)=N_1\left|\omega \beta_{1} A_1(\omega)\right|^2+N_2\left|\omega \beta_{2} A_2(\omega) \right|^2.
\end{aligned}
\label{eq:Iabs}
\end{equation}

\subsection{\label{sec:PL}Photoluminescence}
For the PL spectrum, the derivation is similar to that of Ref. \cite{coupley}, therefore, we would rewrite some of the significance conclusions. The eigen solutions of PL should satisfy the following equations:
\begin{equation}
\begin{aligned}
&\ddot{x}_1+ \beta_{1} \dot{x}_1+\omega_{1}^2 x_1+\eta_{2}\ddot{x}_2+\gamma_{2} \dot{x}_2+g_{2}^2 x_2=0,
\\
&\ddot{x}_2+ \beta_{2} \dot{x}_2+\omega_{2}^2 x_2+\eta_{1}\ddot{x}_1+\gamma_{1} \dot{x}_1+g_{1}^2 x_1=0.
\end{aligned}
\label{eq:couple2}
\end{equation}
Assume the solutions of PL are $x_j(t)=B_j\mathrm{exp}(\alpha t)$ for $j=1,\ 2$, the solutions of $\alpha$ satisfy:
\begin{equation}
\begin{vmatrix}
\Omega_1(\alpha) & G_2(\alpha) \\
G_1(\alpha) & \Omega_2(\alpha)
\end{vmatrix}
=0.
\label{eq:PLm}
\end{equation}
There are generally 4 solutions for $\alpha$. Although the expressions of $\alpha$ are complicated, we could write it in a simple form:
\begin{equation}
\begin{aligned}
\alpha_1=-\frac{\beta_1^{\prime}}{2}-\mathrm{i}\omega_1^{\prime},~~
\alpha_2=-\frac{\beta_1^{\prime}}{2}+\mathrm{i}\omega_1^{\prime},\\
\alpha_3=-\frac{\beta_2^{\prime}}{2}-\mathrm{i}\omega_2^{\prime},~~
\alpha_4=-\frac{\beta_2^{\prime}}{2}+\mathrm{i}\omega_2^{\prime}.
\end{aligned}
\label{eq:alpha}
\end{equation}
Obviously, it appears four modes, but two of them are independent, with two new eigenfrequencies ($\omega_1^{\prime}$ and $\omega_2^{\prime}$) and two new damping coefficients ($\beta_1^{\prime}$ and $\beta_2^{\prime}$).
The analysis of these modes has been discussed in detail in Ref. \cite{coupley}. The total solutions of $x_j(t)$ can be written as:
\begin{equation}
x_j(t)=A_j \mathrm{exp}(\alpha_0 t)+ \sum_{k=1}^{4} B_{jk} \mathrm{exp}(\alpha_k t),~~
\mathrm{for}~j=1,2.
\label{eq:xt}
\end{equation}
Here, $B_{jk}$ refers to the amplitude of the $j$th oscillator in the $k$th mode, and we define $\alpha_0=-\mathrm{i}\omega_{ex}$, then $B_{jk}$ is expressed as:
\begin{equation}
B_{jk}=\frac{-A_j \prod \limits_{n\neq k }^4 (\alpha_0- \alpha_n) -\ddot{x}_j(0)\sum \limits_{n\neq k}^4 \alpha_n +\dddot{x}_j(0)}
{\prod \limits_{n\ne k}^4(\alpha_k-\alpha_n)}.
\label{eq:Bjk}
\end{equation}
Here, the initial conditions are $x_j(0)=0$ and $\dot{x}_j(0)=0$, and $\ddot{x}_j(0)$ and $\dddot{x}_j(0)$ can be derived from Eq. \ref{eq:couple1}.
After considering the angular $\theta_2$, the PL spectrum can be written as \cite{Cheng2018,couplex,coupley}:
\begin{equation}
I_{PL}(\omega)=\sum \limits_{k=1}^4
 \frac{|\mathrm{Re}(\alpha_k)|  B_{k}^{\prime} }{\left( \omega+\mathrm{Im}(\alpha_k) \right)^2+(\mathrm{Re}(\alpha_k))^2}.
\label{eq:IPL}
\end{equation}
Here, $\mathrm{Re}(\alpha_k)$ and $\mathrm{Im}(\alpha_k)$ are the real and imaginary parts of $\alpha_k$, respectively, and $B_{k}^{\prime}=|\alpha_k|^4(\left|N_1 B_{1k}+N_2 B_{2k}\cos \theta_2 \right|^2 + \left| N_2 B_{2k}\sin \theta_2 \right|^2)$ for $k=1,~2,~3,~4$.

\subsection{\label{sec:chirality}Chirality}
To analyze the chirality of the system, we should compare the spectra between the right- and left-handed circularly polarized incident light. Assume $E_x=E_0$, hence, for the RCP one, $E_y=E_0\mathrm{exp}(-\mathrm{i}\pi/2)$, the spectrum is denoted by $I_R$; for the LCP one $E_y=E_0\mathrm{exp}(+\mathrm{i}\pi/2)$, the spectrum is denoted by $I_L$. The intensity and line shape varies with the rotation of the incident light. We define the chirality of the system as:
\begin{equation}
Chirality=\frac{I_R-I_L}{I_R+I_L}.
\label{eq:chirality}
\end{equation}
Therefore, the chirality of the scattering, absorption, and PL spectra can be evaluated.

\section{\label{sec:Results}Results and Discussions}
\begin{figure}[tb]
\includegraphics[width=0.48\textwidth]{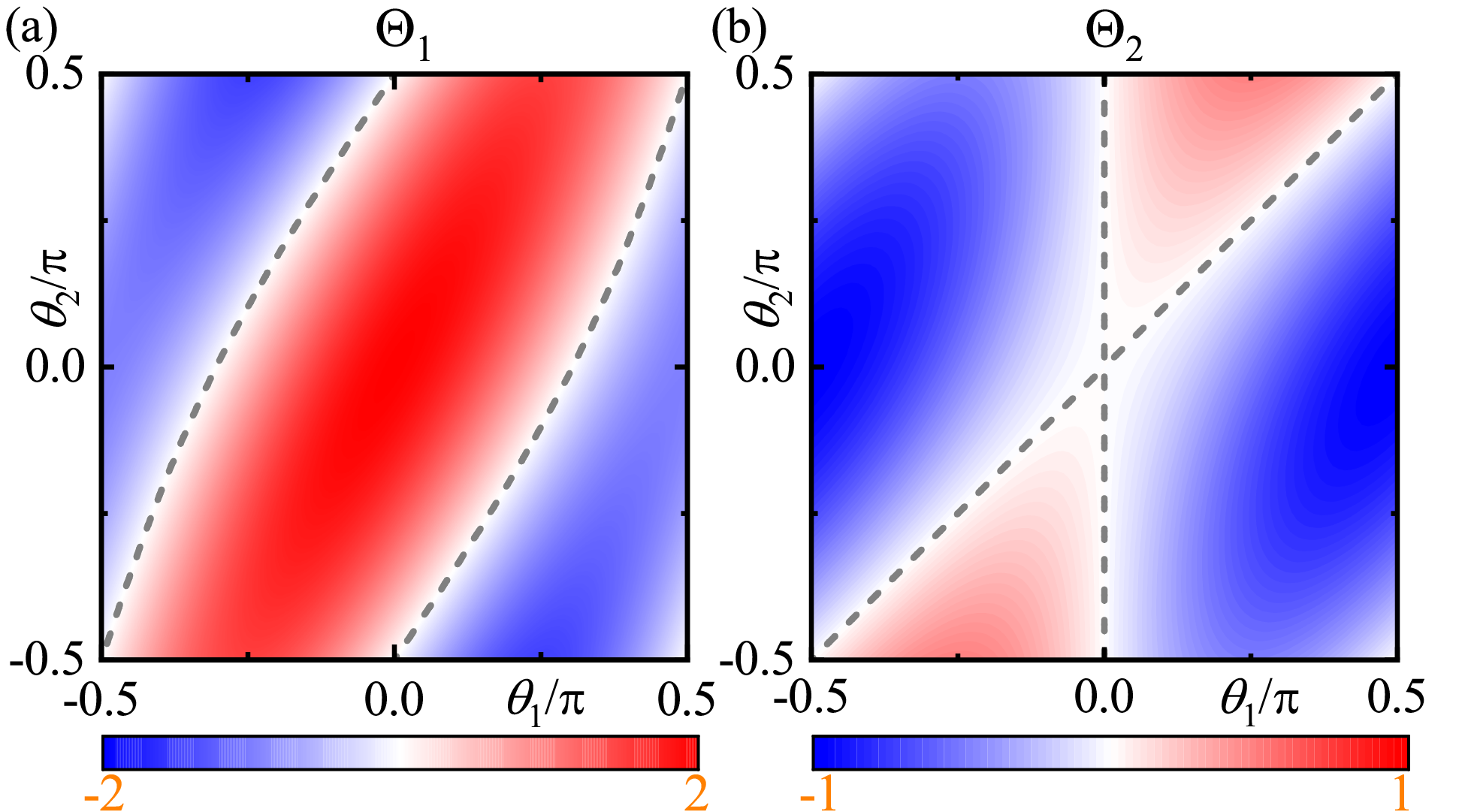}
\caption{\label{fig:Theta} $\Theta_1$ (a) and $\Theta_2$ (b) as a function of $(\theta_1, \theta_2)$, calculated from Eq. \ref{eq:Th12}. The color bars stand for the values of $\Theta_1$ and $\Theta_2$, respectively.
Gray dashed curves represent the contour lines of the value of 0.
}
\end{figure}

From Eq. \ref{eq:CS}, we notice that the coupling coefficients are proportional to $\Theta_1$ or $\Theta_2$, which are significant in the dynamics of the oscillators. Fig. \ref{fig:Theta} shows the values of $\Theta_1(\theta_1,\theta_2)$ and $\Theta_2(\theta_1,\theta_2)$, in the range of $-\pi/2\leq \theta_1,\theta_2 \leq \pi/2$. Obviously, $\Theta_1$ and $\Theta_2$ are centrosymmetric with $(\theta_1,\theta_2)$, and the value ranges are $-1\leq \Theta_1\leq 2$ and $-1\leq \Theta_2\leq 1/2$. $\Theta_1$ reaches its maximum at $(\theta_1,\theta_2)=(0,0)$, and minimum at $(\theta_1,\theta_2)=(\pi/4,-\pi/2)$ and $(-\pi/4,\pi/2)$; $\Theta_2$ reaches its maximum at $(\theta_1,\theta_2)=(\pi/4,\pi/2)$ and $(-\pi/4,-\pi/2)$, and minimum at $(\theta_1,\theta_2)=(\pm \pi/2,0)$. Generally, in a practical case, $|\eta_j| \ll 1$ so that $g_j$ and $\gamma_j$ play dominant roles in the coupling, i.e., $\Theta_1$ is much more significant than $\Theta_2$. Hence, the contour lines of $\Theta_1=0$ indicate that there is (almost) no coupling at these angles, no matter how close the two MNPs are.

\subsection{\label{sec:Results:ScaAbs}Scattering and absorption}
First, we discuss the chirality of the scattering and absorption spectra.

\begin{figure}[tb]
\includegraphics[width=0.48\textwidth]{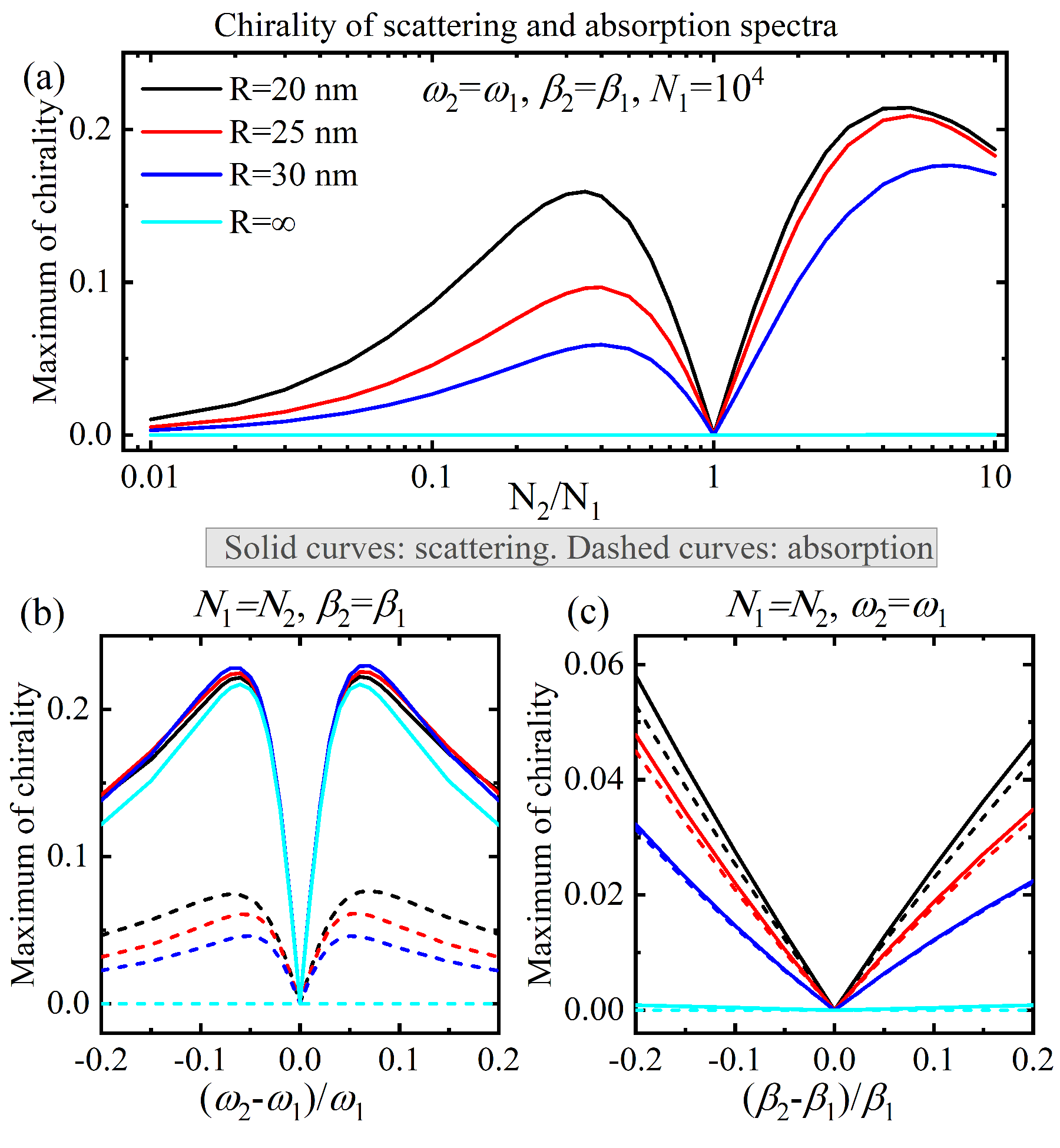}
\caption{\label{fig:ChiScaM} Maximum of the chirality of the scattering and absorption spectra as a function of $N_2$ (a), $\omega_2$ (b), and $\beta_2$ (c).
Solid and red dashed curves stand for the scattering and absorption, respectively. Black, red, blue, and cyan stand for the distance $R=20$ nm, 25 nm, 30 nm, and $\infty$, respectively. Here, $\omega_1=2.069$ eV, $\beta_1=0.1034$ eV, (corresponding to the resonant wavelength of $\lambda_{c}=600$ nm), and $N_1=10^4$.
}
\end{figure}
\begin{figure}[tb]
\includegraphics[width=0.48\textwidth]{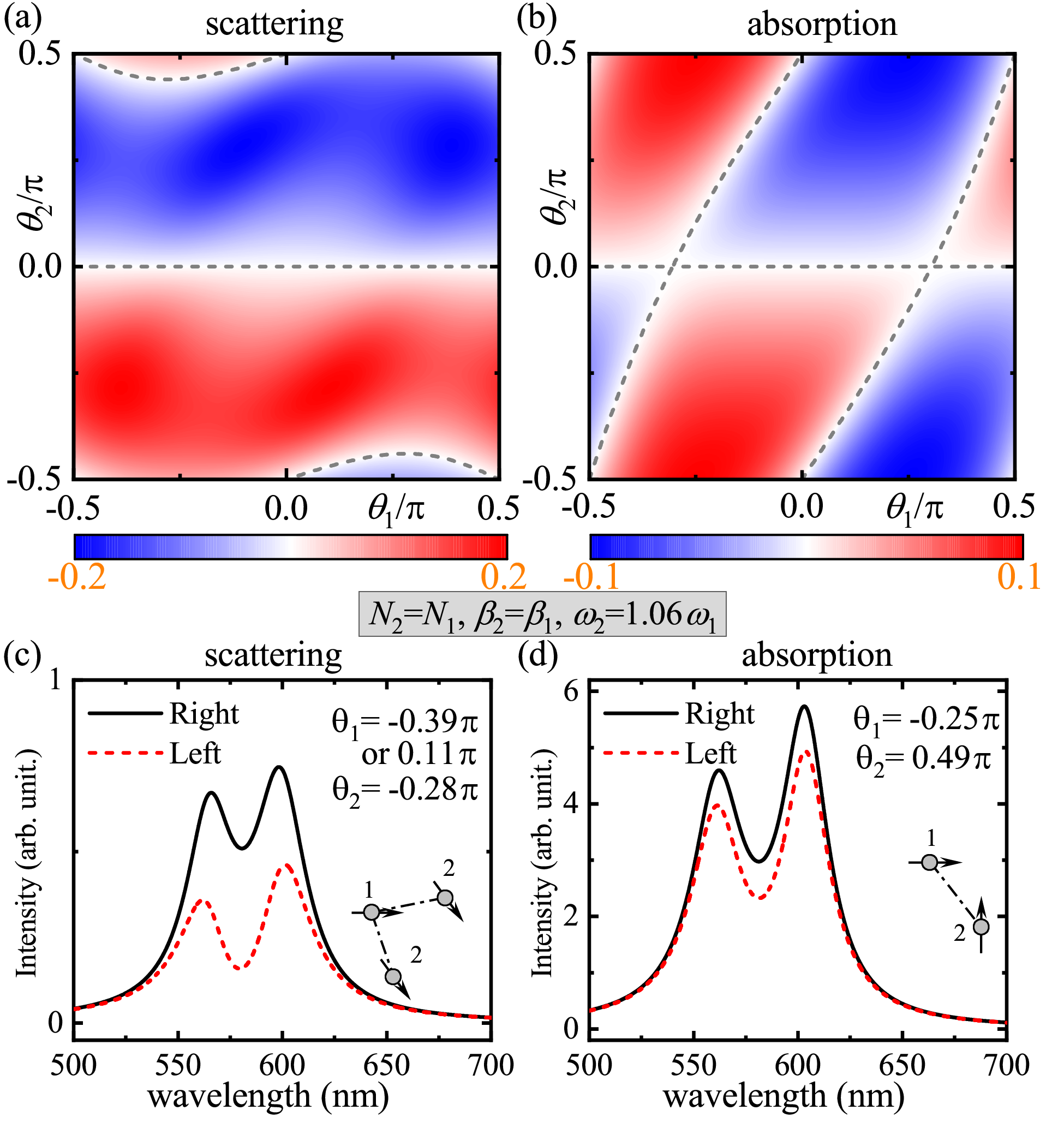}
\caption{\label{fig:ChiSca2} (a) and (b) Chirality of the scattering and absorption spectra as a function of $(\theta_1, \theta_2)$, respectively. Gray dashed curves represent the contour lines of the value of 0.
The color bars stand for the values of corresponding chirality.
(c) and (d) Scattering and absorption spectra excited with RCP (black solid) and LCP (red dashed) light at the maximum of the chirality of (a) and (b), respectively; the angles at which the chirality reaches the maximum are $(\theta_1,\theta_2)=(-0.39\pi~\mathrm{or}~0.11\pi,-0.28\pi)$ (c) and $(\theta_1,\theta_2)=(-0.25\pi,0.49\pi)$ (d), respectively.
The insets of (c) and (d) represent their configurations of the MNPs, respectively.
Here, $R=20$ nm, $N_1=N_2=10^4$, $\omega_1=2.069$ eV, $\omega_2=1.06\omega_1$, and $\beta_1=\beta_2=0.1034$ eV.
}
\end{figure}

Fig. \ref{fig:ChiScaM} shows the maximum of the chirality as a function of $N_2$, $\omega_2$, and $\beta_2$, respectively,  varying with $R$. Here, the maximum is defined as: for a certain $N_2$, $\omega_2$, or $\beta_2$, we calculate the chirality as a function of $(\theta_1,\theta_2)$, in which we find the maximum value.
In, Fig. \ref{fig:ChiScaM}a, the chirality of the absorption spectra is 0 for all the $R$ and $N_2$. However, the chirality of the scattering spectra is not 0. When $N_1=N_2$, the system is symmetric in the parameters of the MNPs, which show no chirality. The chirality appears as $N_1 \neq N_2$, which breaks the symmetry. Evidently, there appears two local maximum values for $N_2<N_1$ and $N_2>N_1$, respectively.
When $N_2<N_1$, as $N_2$ decreases, the breaking of symmetry gets greater; on the other hand, the differences between the two groups of the coupling coefficients, i.e., $\Delta g=|g_{21}-g_{12}|,~\Delta \gamma=|\gamma_{21}-\gamma_{12}|,~$ and $\Delta \eta=|\eta_{21}-\eta_{12}|$, increase. The former increases the chirality, but the latter decreases the chirality. Hence, the first local maximum value (at about $N_2=0.35N_1$ with $R=20$ nm) is derived from both the two factors. The reason of the second one (at about $N_2=5N_1$ with $R=20$ nm) is similar, hence we do not describe here.
As $R$ in creases, i.e., the coupling strength decreases, the chirality decreases. This indicates that in this case (only varying $N_2$) the chirality of the scattering is greatly influenced by the coupling strength, which means this chirality originates from the coupling.
In Fig. \ref{fig:ChiScaM}b, both the scattering and absorption spectra have chirality, and the former is much larger then the latter. Similar to Fig. \ref{fig:ChiScaM}a, there are two local maximum values for $\omega_2<\omega_1$ and $\omega_2>\omega_1$, respectively. The reason is also similar to Fig. \ref{fig:ChiScaM}a. The chirality is sensitive to $\Delta\omega=|\omega_2-\omega_1|$, the maximum values of which are at about $\Delta \omega=0.06\omega_1$.
As the coupling strength decreases, the chirality of the absorption spectra decreases, but the scattering one changes little. This indicates that the chirality of the scattering spectra does not originate from the coupling, the origin of which will be discussed later in Fig. \ref{fig:ChiSca2}.
In Fig. \ref{fig:ChiScaM}c, both the scattering and absorption spectra have chirality with little difference, and the values of the both are so small that we can ignore the chirality of this case. As the coupling strength decreases, both of the chirality decrease. We do not find a local maximum value in this case (only varying $\beta_2$), because the difference $\Delta \beta=|\beta_2-\beta_1|$ is not large enough to introduce the its influence on the chirality, i.e., the breaking of symmetry plays an more important role.

To analyze the chirality and its behavior, some details are included.

\begin{figure}[tb]
\includegraphics[width=0.48\textwidth]{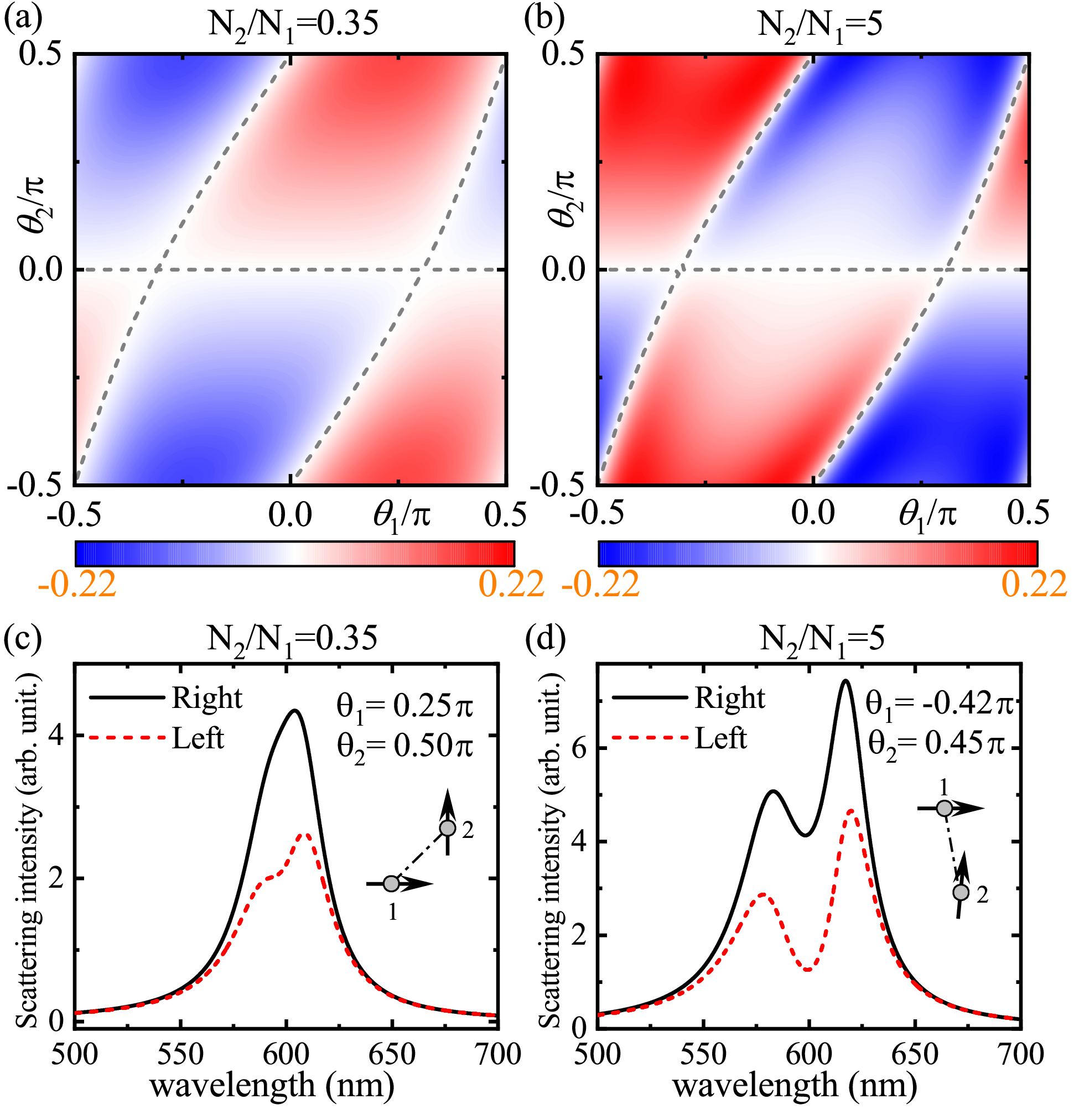}
\caption{\label{fig:ChiSca1} (a) and (b) Chirality of the scattering spectra as a function of $(\theta_1, \theta_2)$ with $N_2/N_1=0.35$ and $N_2/N_1=5$, respectively. Gray dashed curves represent the contour lines of the value of 0.
The color bars stand for the values of corresponding chirality.
(c) and (d) Scattering spectra excited with RCP (black solid) and LCP (red dashed) at the maximum of the chirality of (a) and (b), respectively; the angles at which the chirality reaches the maximum are $(\theta_1,\theta_2)=(0.25\pi,0.50\pi)$ (c) and $(\theta_1,\theta_2)=(-0.42\pi,0.45\pi)$ (d), respectively.
The insets of (c) and (d) represent their configurations of the MNPs, respectively.
Here, $R=20$ nm, $N_1=10^4$, $\omega_1=\omega_2=2.069$ eV, and $\beta_1=\beta_2=0.1034$ eV.
}
\end{figure}

Fig. \ref{fig:ChiSca1}a and \ref{fig:ChiSca1}b show the chirality of the scattering spectra as a function of $(\theta_1,\theta_2)$ with $N_2/N1=0.35$ and $N_2/N_1=5$, respectively, which are corresponding to the two local maximums in Fig. \ref{fig:ChiScaM}a. The contour lines of value 0 (except for the ``$\theta_2=0$'' lines) are similar to the one of $\Theta_1$ in Fig. \ref{fig:Theta}a. It indicates that the chirality of this case is related to the coupling strength, which agrees with the conclusion that it originates from the coupling strength in Fig. \ref{fig:ChiScaM}a. Furthermore, the chirality also originates from the breaking of symmetry, because the chirality gets higher at more nonsymmetric point of $(\theta_1,\theta_2)$.
Fig. \ref{fig:ChiSca1}c and \ref{fig:ChiSca1}d show the scattering spectra excited with RCP and LCP light, corresponding to Fig. \ref{fig:ChiSca1}a and \ref{fig:ChiSca1}b, respectively. Here, we choose the configurations of $(\theta_1,\theta_2)$ at which the chirality reaches their maximums. The results show that the scattering spectra are evidently different between RCP and LCP cases. Due to the strong coupling, the spectra split into to modes, which has been discussed in our previous work in detail \cite{couplex,coupley}. We focus on the chirality properties in this work, hence, we are not going to discuss much of the splitting.

Fig. \ref{fig:ChiSca2}a and \ref{fig:ChiSca2}b show the chirality of the scattering and absorption spectra as a function of $(\theta_1,\theta_2)$, respectively, which are corresponding to the two local maximums in Fig. \ref{fig:ChiScaM}b. In Fig. \ref{fig:ChiSca2}b, the contour lines of value 0 (except for the ``$\theta_2=0$'' lines) are similar to the one of $\Theta_1$ in Fig. \ref{fig:Theta}a, which indicates the origin of the chirality of the absorption spectra is related to the coupling strength, which also agrees with the conclusion that it originates from the coupling strength in Fig. \ref{fig:ChiScaM}b. However, in Fig. \ref{fig:ChiSca2}a, the ``0'' contour lines are different from the one of $\Theta_1$ in Fig. \ref{fig:Theta}a. It indicates that in this case (only varying $\omega_2$) the coupling plays an unimportant role in the chirality of the scattering spectra. This agrees with the behaviour of the chirality in Fig. \ref{fig:ChiScaM}b (solid lines). Hence, we can summarize that in this case the breaking of symmetry plays a more important role in the chirality of the scattering spectra.
Fig. \ref{fig:ChiSca2}c and \ref{fig:ChiSca2}d show the scattering and absorption spectra excited with RCP and LCP light, corresponding to Fig. \ref{fig:ChiSca2}a and \ref{fig:ChiSca2}b, respectively. Here, we choose the configurations of $(\theta_1,\theta_2)$ at which the
chirality reaches their maximums. The results show that
the scattering spectra are evidently different between
RCP and LCP cases, but the absorption ones are not. The reason can be found in Eq. \ref{eq:Isca} and Eq. \ref{eq:Iabs}. In the scattering, the spectra are from the coherent superposition of the emitted electric field, i.e., the formula has the cross terms (coherent terms) which are related to $(\theta_1,\theta_2)$ that are relevant to the symmetry, and contribute to the chirality; while in the absorption, the spectra are the addition of the respective intensity, and has no cross terms and no relation to $(\theta_1,\theta_2)$. The only contribution of the chirality of the absorption spectra is the coupling strength which is related to $(\theta_1,\theta_2)$. Therefore, the chirality of the scattering is larger than the one of the absorption.

\subsection{\label{sec:results:PL}Photoluminescence}
Second, we discuss the chirality of the PL spectra.

\begin{figure}[tb]
\includegraphics[width=0.48\textwidth]{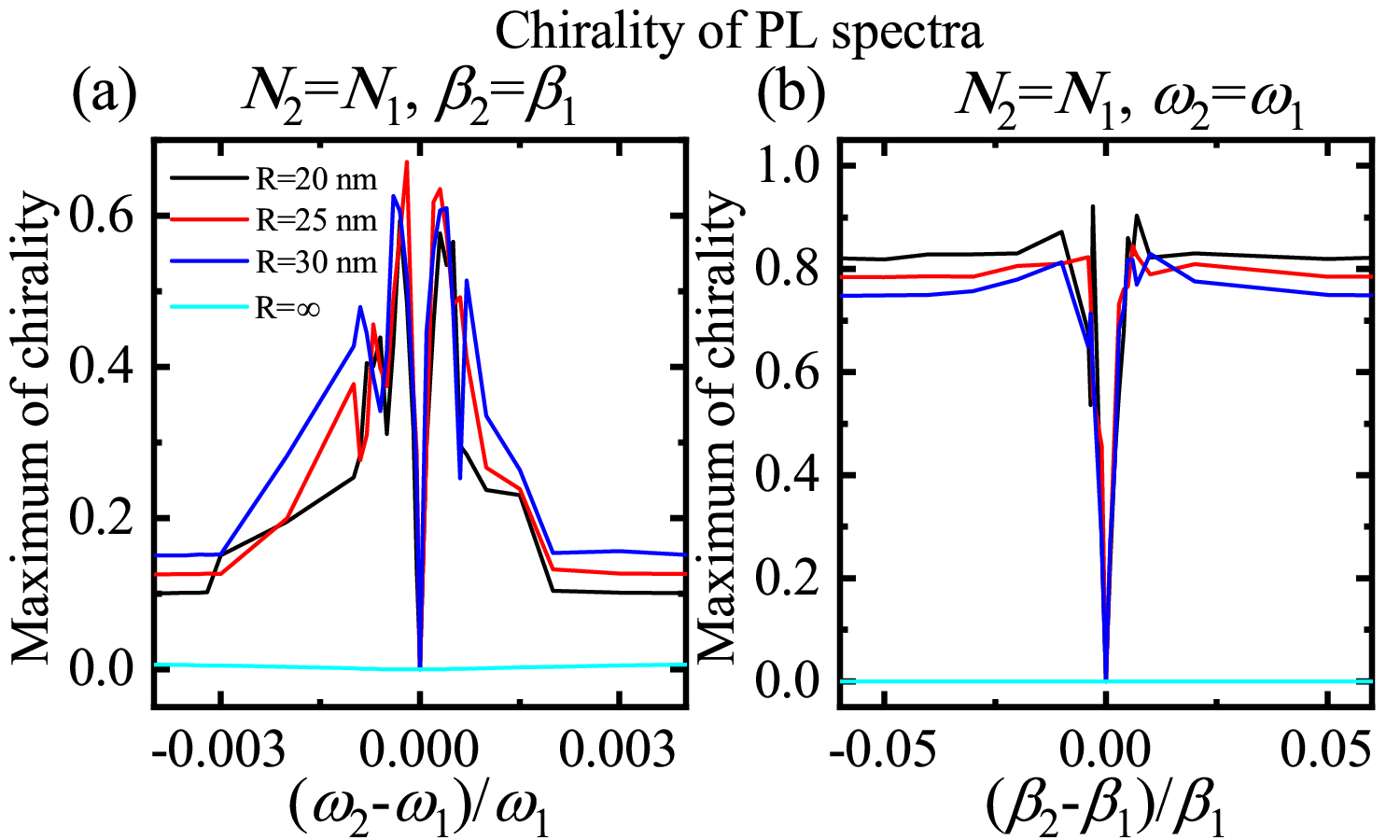}
\caption{\label{fig:ChiPLM} Maximum of the chirality of the PL spectra as a function of $\omega_2$ (a), and $\beta_2$ (b).
Black, red, blue, and cyan curves stand for the distance of $R=20$ nm, 25 nm, 30 nm, and $\infty$, respectively.
Here, $\omega_1=2.069$ eV, $\beta_1=0.1034$ eV, and $N_1=N_2=10^4$.
}
\end{figure}

Fig. \ref{fig:ChiPLM} shows the maximum of the chirality as a function of $\omega_2$ and $\beta_2$, respectively, varying with $R$.
In Fig. \ref{fig:ChiPLM}a, when $|\Delta \omega|$ is very small, the chirality reaches a maximum of about 0.6; when $|\Delta \omega|$ gets larger, the chirality decreases to about 0.1-0.2.
In Fig. \ref{fig:ChiPLM}b, when $|\Delta \beta|$ increases from 0, the chirality increases rapidly from 0 to about 0.8.
For both cases, in strong coupling, the chirality varies little with $R$; when there is no coupling, the chirality becomes 0. This consequence indicates that $\Delta \beta$ affects the chirality of the PL spectra more than $\Delta \omega$ does; the chirality can be quite sensitive to a small value of $\Delta \omega$ and $\Delta \beta$.
We also calculate the chirality as a function of $N_2$ with $\omega_1=\omega_2$ and $\beta_1=\beta_2$, but the value is almost 0. It indicates that in this case (varying $N_2$), there is no chirality in the system, hence we do not show it in the figure.

To analyze the chirality and its behavior, some details
are included.

\begin{figure}[tb]
\includegraphics[width=0.48\textwidth]{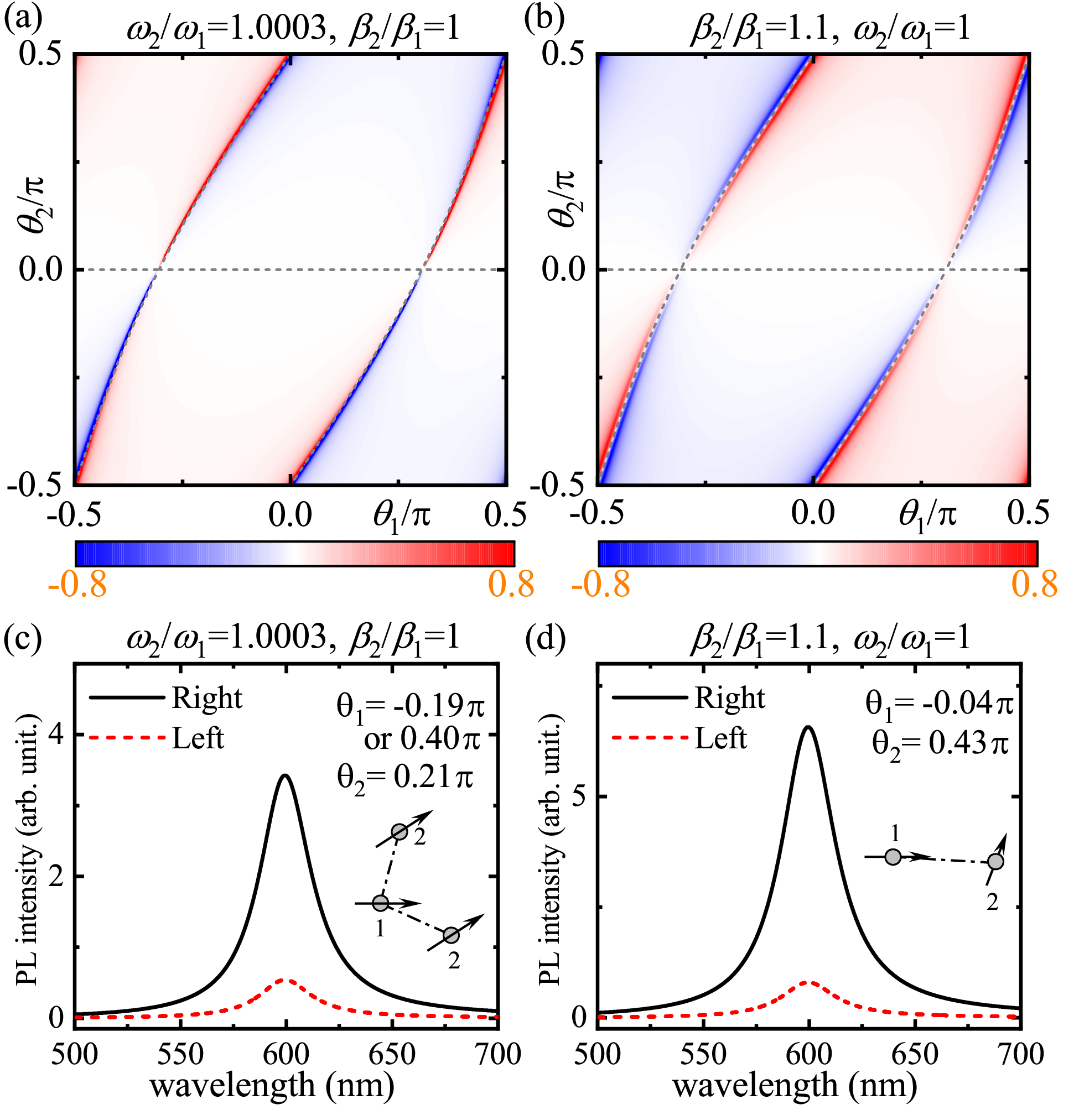}
\caption{\label{fig:ChiPL} (a) and (b) Chirality of the PL spectra as a function of $(\theta_1, \theta_2)$ with $\omega_2=1.0003\omega_1$ and $\beta_2=1.1\beta_1$, respectively.
Gray dashed curves represent the contour lines of the value of 0. The color bars stand for the values of corresponding chirality.
(c) and (d) PL spectra excited with RCP (black solid) and LCP (red dashed) light at the maximum of the chirality of (a) and (b), respectively; the angles at which the chirality reaches the maximum are $(\theta_1,\theta_2)=(-0.19\pi~\mathrm{or}~0.40\pi,0.21\pi)$ (c) and $(\theta_1,\theta_2)=(-0.04\pi,0.43\pi)$ (d), respectively.
The insets of (c) and (d) represent their configurations of the MNPs, respectively.
Here, $R=20$ nm, $N_1=N_2=10^4$, $\omega_1=2.069$ eV, and $\beta_1=0.1034$ eV.
}
\end{figure}

Fig. \ref{fig:ChiPL}a and \ref{fig:ChiPL}b show the chirality of the PL spectra as a function of $(\theta_1,\theta_2)$ with $\omega_2/\omega_1=1.0003$ and $\beta_2=\beta_1$, respectively. The former is corresponding to the maximum in Fig. \ref{fig:ChiPLM}a ($R=20$ nm), and the latter is corresponding to a general case in Fig.  \ref{fig:ChiPLM}b.
The contour lines of value 0 (except for the ``$\theta_2=0$'' lines) are similar to the one of $\Theta_1$ in Fig. \ref{fig:Theta}a. Obviously, the pattern is different from Fig. \ref{fig:ChiSca1} and \ref{fig:ChiSca2}. For the scattering, the maximum appears usually at the point far from the 0 contour lines; however, for the PL, the maximum appears near the 0 contour lines. This phenomenon indicates that the chirality of the PL spectra is quite sensitive to $(\theta_1,\theta_2)$ near the contour lines, the sensitivity of which is similar to that of $\Delta \omega$ and $\Delta \beta$.
Fig. \ref{fig:ChiPL}c and \ref{fig:ChiPL}d show the PL spectra excited with RCP and LCP light, corresponding to  Fig. \ref{fig:ChiPL}a and \ref{fig:ChiPL}b, respectively. Here, we choose the configurations of $(\theta_1,\theta_2)$ at which the chirality reaches their maximums. The results show that the PL spectra are different between RCP and LCP cases. Although $R=20$ nm corresponds to the strong coupling case, there is no splitting  for the maximum configurations. Because, as mentioned above, the chirality reaches the maximum near the 0 contour lines, resulting in the negligible coupling strength at these $(\theta_1,\theta_2)$, the spectra show no splitting.

\subsection{\label{sec:results:Control}Controlling spectra}
\begin{figure}[tb]
\includegraphics[width=0.48\textwidth]{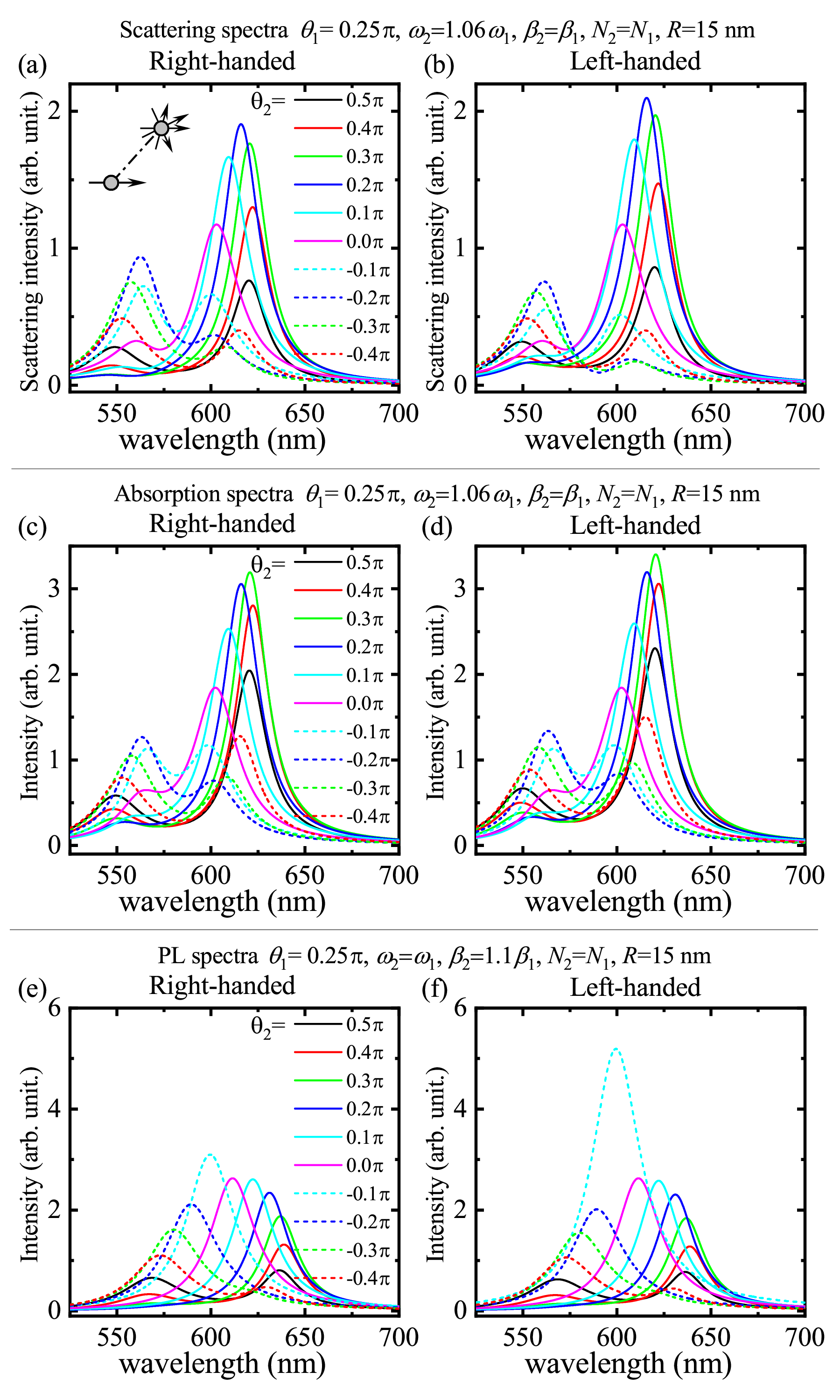}
\caption{\label{fig:ScaAbsPL} Scattering (a-b), absorption (c-d), and PL (e-f) spectra varying with $\theta_2$. (a), (c), and (e) represent RCP external light; (b), (d), and (f) represent LCP external light.
Here, $R=15$ nm, $\theta_1=0.25\pi$, $\omega_1=2.069$ eV, $\beta_1=0.1034$ eV, and $N_1=N_2=10^4$
}
\end{figure}
Last, we show a way to control the spectra of the system, i.e., changing the configurations of the two MNPs.

Fig. \ref{fig:ScaAbsPL} shows an example of a general case of the scattering, absorption, and PL spectra excited with RCP and LCP light, at $\theta_1=\pi/4$ varying with $\theta_2$. As $\theta_2$ decreases from $\pi/2$ to $-\pi/2$, the resonance frequencies of the red mode ($\lambda_c>600$ nm) increases and then decreases accompany with the change of the intensities; the ones of the blue mode ($\lambda_c<600$ nm) behave oppositely. If we conect all the peak points of the red (blue) mode in sequence, we can obtain a left-handed (right-handed) circle (not shown), which is quite interesting. Therefore, we can conclude that, the resonance frequencies, damping coefficients, and intensities of all the spectra can be controlled by $\theta_2$. The difference between RCP and LCP cases (the chirality) depends on the position ($\theta_1,\theta_2$) at which the configurations are. Therefore, the chirality is also controlled by $\theta_2$.

\section*{\label{sec:Conclusion}Conclusions}
In conclusion,
there are three important factors which directly influence the chirality of the coupled MNPs: the symmetry(factor 1), the coupling strength (factor 2), and the coherent superposition of the electric field (factor 3). Strictly speaking, factor 2 and 3 can be influenced by factor 1, but we separate them to illustrate the mechanisms of the chirality of different spectra.
Among them, factors 1-3 are influenced by the orientations of the MNPs, i.e., $(\theta_1,\theta_2)$, and factor 1 is additionally influenced by the difference between the MNPs' eigen parameters (parameter-difference), i.e., $\Delta N$, $\Delta \omega$, and $\Delta \beta$. These three factors play different proportional roles in the chirality of the scattering, absorption, and PL spectra.
In particular, for the scattering, when $\Delta \omega=\Delta \beta=0$, factors 1-3 are all important; when $\Delta N=0,~\Delta \beta=0$, factor 2 is not important compared with factors 1 and 3. The chirality is not that large (about 0.2) even in strong coupling.
For the absorption, factors 1 and 2 play important roles, but the chirality is small (less than 0.1).
For the PL, when  $\Delta \omega=\Delta \beta=0$, no chirality is obtained; when $\Delta N=0,~\Delta \beta=0$, the chirality is extremely sensitive to $\omega_2$; when $\Delta N=0,~\Delta \omega=0$, the chirality gets large (about 0.8) with a small $\Delta \beta$.
Moreover, all the spectra can be controlled by varying the orientations $(\theta_1,\theta_2)$.
This work provides a deeper understanding on the chiral plasmonics and may guide relevant applications in theory.

\section*{\label{sec:Acknowldegment}Acknowledgment}
This work was supported by the Fundamental Research Funds for the Central Universities (Grant No. FRF-TP-20-075A1).

\section*{Disclosures}
The authors declare no conflicts of interest.

\section*{Data availability}
The data that support the findings of this study are available from the corresponding author upon reasonable request.

\section*{\label{sec:Ref}References}

\bibliography{CM_Cheng}

\begin{thebibliography}{25}%
\makeatletter
\providecommand \@ifxundefined [1]{%
 \@ifx{#1\undefined}
}%
\providecommand \@ifnum [1]{%
 \ifnum #1\expandafter \@firstoftwo
 \else \expandafter \@secondoftwo
 \fi
}%
\providecommand \@ifx [1]{%
 \ifx #1\expandafter \@firstoftwo
 \else \expandafter \@secondoftwo
 \fi
}%
\providecommand \natexlab [1]{#1}%
\providecommand \enquote  [1]{``#1''}%
\providecommand \bibnamefont  [1]{#1}%
\providecommand \bibfnamefont [1]{#1}%
\providecommand \citenamefont [1]{#1}%
\providecommand \href@noop [0]{\@secondoftwo}%
\providecommand \href [0]{\begingroup \@sanitize@url \@href}%
\providecommand \@href[1]{\@@startlink{#1}\@@href}%
\providecommand \@@href[1]{\endgroup#1\@@endlink}%
\providecommand \@sanitize@url [0]{\catcode `\\12\catcode `\$12\catcode
  `\&12\catcode `\#12\catcode `\^12\catcode `\_12\catcode `\%12\relax}%
\providecommand \@@startlink[1]{}%
\providecommand \@@endlink[0]{}%
\providecommand \url  [0]{\begingroup\@sanitize@url \@url }%
\providecommand \@url [1]{\endgroup\@href {#1}{\urlprefix }}%
\providecommand \urlprefix  [0]{URL }%
\providecommand \Eprint [0]{\href }%
\providecommand \doibase [0]{http://dx.doi.org/}%
\providecommand \selectlanguage [0]{\@gobble}%
\providecommand \bibinfo  [0]{\@secondoftwo}%
\providecommand \bibfield  [0]{\@secondoftwo}%
\providecommand \translation [1]{[#1]}%
\providecommand \BibitemOpen [0]{}%
\providecommand \bibitemStop [0]{}%
\providecommand \bibitemNoStop [0]{.\EOS\space}%
\providecommand \EOS [0]{\spacefactor3000\relax}%
\providecommand \BibitemShut  [1]{\csname bibitem#1\endcsname}%
\let\auto@bib@innerbib\@empty
\bibitem [{\citenamefont {Prelog}(1976)}]{chirality}%
  \BibitemOpen
  \bibfield  {author} {\bibinfo {author} {\bibfnamefont {V.}~\bibnamefont
  {Prelog}},\ }\href {\doibase 10.1126/science.935852} {\bibfield  {journal}
  {\bibinfo  {journal} {Science}\ }\textbf {\bibinfo {volume} {193}},\ \bibinfo
  {pages} {17} (\bibinfo {year} {1976})}\BibitemShut {NoStop}%
\bibitem [{\citenamefont {Wu}\ and\ \citenamefont {Pauly}(2022)}]{Chirality2}%
  \BibitemOpen
  \bibfield  {author} {\bibinfo {author} {\bibfnamefont {W.}~\bibnamefont
  {Wu}}\ and\ \bibinfo {author} {\bibfnamefont {M.}~\bibnamefont {Pauly}},\
  }\href {\doibase 10.1039/D1MA00915J} {\bibfield  {journal} {\bibinfo
  {journal} {Materials Advances}\ }\textbf {\bibinfo {volume} {3}},\ \bibinfo
  {pages} {186} (\bibinfo {year} {2022})}\BibitemShut {NoStop}%
\bibitem [{\citenamefont {Chen}\ \emph {et~al.}(2021)\citenamefont {Chen},
  \citenamefont {Gao}, \citenamefont {Zheng}, \citenamefont {Liu},
  \citenamefont {Meng}, \citenamefont {Li}, \citenamefont {Cai}, \citenamefont
  {Fan}, \citenamefont {Ji},\ and\ \citenamefont {Wu}}]{appLC1}%
  \BibitemOpen
  \bibfield  {author} {\bibinfo {author} {\bibfnamefont {J.}~\bibnamefont
  {Chen}}, \bibinfo {author} {\bibfnamefont {X.}~\bibnamefont {Gao}}, \bibinfo
  {author} {\bibfnamefont {Q.}~\bibnamefont {Zheng}}, \bibinfo {author}
  {\bibfnamefont {J.}~\bibnamefont {Liu}}, \bibinfo {author} {\bibfnamefont
  {D.}~\bibnamefont {Meng}}, \bibinfo {author} {\bibfnamefont {H.}~\bibnamefont
  {Li}}, \bibinfo {author} {\bibfnamefont {R.}~\bibnamefont {Cai}}, \bibinfo
  {author} {\bibfnamefont {H.}~\bibnamefont {Fan}}, \bibinfo {author}
  {\bibfnamefont {Y.}~\bibnamefont {Ji}}, \ and\ \bibinfo {author}
  {\bibfnamefont {X.}~\bibnamefont {Wu}},\ }\href {\doibase
  10.1021/acsnano.1c05489} {\bibfield  {journal} {\bibinfo  {journal} {ACS
  Nano}\ }\textbf {\bibinfo {volume} {15}},\ \bibinfo {pages} {15114} (\bibinfo
  {year} {2021})}\BibitemShut {NoStop}%
\bibitem [{\citenamefont {Cheng}\ \emph {et~al.}(2023)\citenamefont {Cheng},
  \citenamefont {Liang}, \citenamefont {Deng}, \citenamefont {Jin},
  \citenamefont {Shangguan}, \citenamefont {Zhang}, \citenamefont {Guo},\ and\
  \citenamefont {Yu}}]{appLC2}%
  \BibitemOpen
  \bibfield  {author} {\bibinfo {author} {\bibfnamefont {H.}~\bibnamefont
  {Cheng}}, \bibinfo {author} {\bibfnamefont {K.}~\bibnamefont {Liang}},
  \bibinfo {author} {\bibfnamefont {X.}~\bibnamefont {Deng}}, \bibinfo {author}
  {\bibfnamefont {L.}~\bibnamefont {Jin}}, \bibinfo {author} {\bibfnamefont
  {J.}~\bibnamefont {Shangguan}}, \bibinfo {author} {\bibfnamefont
  {J.}~\bibnamefont {Zhang}}, \bibinfo {author} {\bibfnamefont
  {J.}~\bibnamefont {Guo}}, \ and\ \bibinfo {author} {\bibfnamefont
  {L.}~\bibnamefont {Yu}},\ }\href {\doibase 10.3390/photonics10030251}
  {\bibfield  {journal} {\bibinfo  {journal} {Photonics}\ }\textbf {\bibinfo
  {volume} {10}},\ \bibinfo {pages} {251} (\bibinfo {year} {2023})}\BibitemShut
  {NoStop}%
\bibitem [{\citenamefont {He}\ \emph {et~al.}(2022)\citenamefont {He},
  \citenamefont {Cen}, \citenamefont {Wang}, \citenamefont {Xu}, \citenamefont
  {Liu}, \citenamefont {Cai}, \citenamefont {Kong}, \citenamefont {Li},
  \citenamefont {Luo}, \citenamefont {Cao},\ and\ \citenamefont
  {Liu}}]{appLC3}%
  \BibitemOpen
  \bibfield  {author} {\bibinfo {author} {\bibfnamefont {H.}~\bibnamefont
  {He}}, \bibinfo {author} {\bibfnamefont {M.}~\bibnamefont {Cen}}, \bibinfo
  {author} {\bibfnamefont {J.}~\bibnamefont {Wang}}, \bibinfo {author}
  {\bibfnamefont {Y.}~\bibnamefont {Xu}}, \bibinfo {author} {\bibfnamefont
  {J.}~\bibnamefont {Liu}}, \bibinfo {author} {\bibfnamefont {W.}~\bibnamefont
  {Cai}}, \bibinfo {author} {\bibfnamefont {D.}~\bibnamefont {Kong}}, \bibinfo
  {author} {\bibfnamefont {K.}~\bibnamefont {Li}}, \bibinfo {author}
  {\bibfnamefont {D.}~\bibnamefont {Luo}}, \bibinfo {author} {\bibfnamefont
  {T.}~\bibnamefont {Cao}}, \ and\ \bibinfo {author} {\bibfnamefont {Y.~J.}\
  \bibnamefont {Liu}},\ }\href {\doibase 10.1021/acsami.2c13267} {\bibfield
  {journal} {\bibinfo  {journal} {ACS Applied Materials \& Interfaces}\
  }\textbf {\bibinfo {volume} {14}},\ \bibinfo {pages} {53981} (\bibinfo {year}
  {2022})}\BibitemShut {NoStop}%
\bibitem [{\citenamefont {Zhang}\ \emph {et~al.}(2023)\citenamefont {Zhang},
  \citenamefont {Shen}, \citenamefont {Gao}, \citenamefont {Peng},
  \citenamefont {Cao}, \citenamefont {Wang}, \citenamefont {Wang},
  \citenamefont {Zhang}, \citenamefont {Yang}, \citenamefont {Liu},\ and\
  \citenamefont {Sun}}]{appLC4}%
  \BibitemOpen
  \bibfield  {author} {\bibinfo {author} {\bibfnamefont {N.-N.}\ \bibnamefont
  {Zhang}}, \bibinfo {author} {\bibfnamefont {Z.-L.}\ \bibnamefont {Shen}},
  \bibinfo {author} {\bibfnamefont {S.-Y.}\ \bibnamefont {Gao}}, \bibinfo
  {author} {\bibfnamefont {F.}~\bibnamefont {Peng}}, \bibinfo {author}
  {\bibfnamefont {Z.-J.}\ \bibnamefont {Cao}}, \bibinfo {author} {\bibfnamefont
  {Y.}~\bibnamefont {Wang}}, \bibinfo {author} {\bibfnamefont {Z.}~\bibnamefont
  {Wang}}, \bibinfo {author} {\bibfnamefont {W.}~\bibnamefont {Zhang}},
  \bibinfo {author} {\bibfnamefont {Y.}~\bibnamefont {Yang}}, \bibinfo {author}
  {\bibfnamefont {K.}~\bibnamefont {Liu}}, \ and\ \bibinfo {author}
  {\bibfnamefont {T.}~\bibnamefont {Sun}},\ }\href {\doibase
  https://doi.org/10.1002/adom.202203119} {\bibfield  {journal} {\bibinfo
  {journal} {Advanced Optical Materials}\ ,\ \bibinfo {pages} {2203119}}
  (\bibinfo {year} {2023})}\BibitemShut {NoStop}%
\bibitem [{\citenamefont {Meng}\ \emph {et~al.}(2022)\citenamefont {Meng},
  \citenamefont {Zhang}, \citenamefont {Liu}, \citenamefont {Li}, \citenamefont
  {Sun}, \citenamefont {Lai},\ and\ \citenamefont {Yu}}]{appDevice1}%
  \BibitemOpen
  \bibfield  {author} {\bibinfo {author} {\bibfnamefont {J.}~\bibnamefont
  {Meng}}, \bibinfo {author} {\bibfnamefont {Z.}~\bibnamefont {Zhang}},
  \bibinfo {author} {\bibfnamefont {W.}~\bibnamefont {Liu}}, \bibinfo {author}
  {\bibfnamefont {Y.}~\bibnamefont {Li}}, \bibinfo {author} {\bibfnamefont
  {Y.}~\bibnamefont {Sun}}, \bibinfo {author} {\bibfnamefont {Z.}~\bibnamefont
  {Lai}}, \ and\ \bibinfo {author} {\bibfnamefont {T.}~\bibnamefont {Yu}},\
  }\href {\doibase 10.1364/OL.472717} {\bibfield  {journal} {\bibinfo
  {journal} {Optics Letters}\ }\textbf {\bibinfo {volume} {47}},\ \bibinfo
  {pages} {5385} (\bibinfo {year} {2022})}\BibitemShut {NoStop}%
\bibitem [{\citenamefont {Zou}\ and\ \citenamefont {Nash}(2022)}]{appDevice2}%
  \BibitemOpen
  \bibfield  {author} {\bibinfo {author} {\bibfnamefont {H.}~\bibnamefont
  {Zou}}\ and\ \bibinfo {author} {\bibfnamefont {G.~R.}\ \bibnamefont {Nash}},\
  }\href {\doibase 10.1364/OME.473926} {\bibfield  {journal} {\bibinfo
  {journal} {Optical Materials Express}\ }\textbf {\bibinfo {volume} {12}},\
  \bibinfo {pages} {4565} (\bibinfo {year} {2022})}\BibitemShut {NoStop}%
\bibitem [{\citenamefont {Bai}\ and\ \citenamefont {Yao}(2021)}]{appDevice3}%
  \BibitemOpen
  \bibfield  {author} {\bibinfo {author} {\bibfnamefont {J.}~\bibnamefont
  {Bai}}\ and\ \bibinfo {author} {\bibfnamefont {Y.}~\bibnamefont {Yao}},\
  }\href {\doibase 10.1021/acsnano.1c02278} {\bibfield  {journal} {\bibinfo
  {journal} {ACS Nano}\ }\textbf {\bibinfo {volume} {15}},\ \bibinfo {pages}
  {14263} (\bibinfo {year} {2021})}\BibitemShut {NoStop}%
\bibitem [{\citenamefont {Zhang}\ \emph {et~al.}(2022)\citenamefont {Zhang},
  \citenamefont {Liu}, \citenamefont {Tian},\ and\ \citenamefont
  {Zhu}}]{appDevice4}%
  \BibitemOpen
  \bibfield  {author} {\bibinfo {author} {\bibfnamefont {F.}~\bibnamefont
  {Zhang}}, \bibinfo {author} {\bibfnamefont {B.}~\bibnamefont {Liu}}, \bibinfo
  {author} {\bibfnamefont {Z.}~\bibnamefont {Tian}}, \ and\ \bibinfo {author}
  {\bibfnamefont {N.}~\bibnamefont {Zhu}},\ }\href {\doibase
  10.35848/1882-0786/ac9a9c} {\bibfield  {journal} {\bibinfo  {journal}
  {Applied Physics Express}\ }\textbf {\bibinfo {volume} {15}},\ \bibinfo
  {pages} {112006} (\bibinfo {year} {2022})}\BibitemShut {NoStop}%
\bibitem [{\citenamefont {Cai}\ \emph {et~al.}(2022)\citenamefont {Cai},
  \citenamefont {Zhang}, \citenamefont {Xu}, \citenamefont {Hao}, \citenamefont
  {Ma}, \citenamefont {Sun}, \citenamefont {Wu}, \citenamefont {Qin},
  \citenamefont {Colombari}, \citenamefont {de~Moura}, \citenamefont {Xu},
  \citenamefont {Silva}, \citenamefont {Carneiro-Neto}, \citenamefont {Gomes},
  \citenamefont {Vall\'{e}e}, \citenamefont {Pereira}, \citenamefont {Liu},
  \citenamefont {Xu}, \citenamefont {Klajn}, \citenamefont {Kotov},\ and\
  \citenamefont {Kuang}}]{appDevice5}%
  \BibitemOpen
  \bibfield  {author} {\bibinfo {author} {\bibfnamefont {J.}~\bibnamefont
  {Cai}}, \bibinfo {author} {\bibfnamefont {W.}~\bibnamefont {Zhang}}, \bibinfo
  {author} {\bibfnamefont {L.}~\bibnamefont {Xu}}, \bibinfo {author}
  {\bibfnamefont {C.}~\bibnamefont {Hao}}, \bibinfo {author} {\bibfnamefont
  {W.}~\bibnamefont {Ma}}, \bibinfo {author} {\bibfnamefont {M.}~\bibnamefont
  {Sun}}, \bibinfo {author} {\bibfnamefont {X.}~\bibnamefont {Wu}}, \bibinfo
  {author} {\bibfnamefont {X.}~\bibnamefont {Qin}}, \bibinfo {author}
  {\bibfnamefont {F.~M.}\ \bibnamefont {Colombari}}, \bibinfo {author}
  {\bibfnamefont {A.~F.}\ \bibnamefont {de~Moura}}, \bibinfo {author}
  {\bibfnamefont {J.}~\bibnamefont {Xu}}, \bibinfo {author} {\bibfnamefont
  {M.~C.}\ \bibnamefont {Silva}}, \bibinfo {author} {\bibfnamefont {E.~B.}\
  \bibnamefont {Carneiro-Neto}}, \bibinfo {author} {\bibfnamefont {W.~R.}\
  \bibnamefont {Gomes}}, \bibinfo {author} {\bibfnamefont {R.~A.~L.}\
  \bibnamefont {Vall\'{e}e}}, \bibinfo {author} {\bibfnamefont {E.~C.}\
  \bibnamefont {Pereira}}, \bibinfo {author} {\bibfnamefont {X.}~\bibnamefont
  {Liu}}, \bibinfo {author} {\bibfnamefont {C.}~\bibnamefont {Xu}}, \bibinfo
  {author} {\bibfnamefont {R.}~\bibnamefont {Klajn}}, \bibinfo {author}
  {\bibfnamefont {N.~A.}\ \bibnamefont {Kotov}}, \ and\ \bibinfo {author}
  {\bibfnamefont {H.}~\bibnamefont {Kuang}},\ }\href {\doibase
  10.1038/s41565-022-01079-3} {\bibfield  {journal} {\bibinfo  {journal}
  {Nature Nanotechnology}\ }\textbf {\bibinfo {volume} {17}},\ \bibinfo {pages}
  {408} (\bibinfo {year} {2022})}\BibitemShut {NoStop}%
\bibitem [{\citenamefont {Lor\'en}\ \emph {et~al.}(2023)\citenamefont
  {Lor\'en}, \citenamefont {Paravicini-Bagliani}, \citenamefont {Saha},
  \citenamefont {Gautier}, \citenamefont {Li}, \citenamefont {Genet},\ and\
  \citenamefont {Mart\'{\i}n-Moreno}}]{appDevice6}%
  \BibitemOpen
  \bibfield  {author} {\bibinfo {author} {\bibfnamefont {F.}~\bibnamefont
  {Lor\'en}}, \bibinfo {author} {\bibfnamefont {G.~L.}\ \bibnamefont
  {Paravicini-Bagliani}}, \bibinfo {author} {\bibfnamefont {S.}~\bibnamefont
  {Saha}}, \bibinfo {author} {\bibfnamefont {J.}~\bibnamefont {Gautier}},
  \bibinfo {author} {\bibfnamefont {M.}~\bibnamefont {Li}}, \bibinfo {author}
  {\bibfnamefont {C.}~\bibnamefont {Genet}}, \ and\ \bibinfo {author}
  {\bibfnamefont {L.}~\bibnamefont {Mart\'{\i}n-Moreno}},\ }\href {\doibase
  10.1103/PhysRevB.107.165128} {\bibfield  {journal} {\bibinfo  {journal}
  {Physical Review B}\ }\textbf {\bibinfo {volume} {107}},\ \bibinfo {pages}
  {165128} (\bibinfo {year} {2023})}\BibitemShut {NoStop}%
\bibitem [{\citenamefont {Maoz}\ \emph {et~al.}(2013)\citenamefont {Maoz},
  \citenamefont {Chaikin}, \citenamefont {Tesler}, \citenamefont {Bar~Elli},
  \citenamefont {Fan}, \citenamefont {Govorov},\ and\ \citenamefont
  {Markovich}}]{appMS1}%
  \BibitemOpen
  \bibfield  {author} {\bibinfo {author} {\bibfnamefont {B.~M.}\ \bibnamefont
  {Maoz}}, \bibinfo {author} {\bibfnamefont {Y.}~\bibnamefont {Chaikin}},
  \bibinfo {author} {\bibfnamefont {A.~B.}\ \bibnamefont {Tesler}}, \bibinfo
  {author} {\bibfnamefont {O.}~\bibnamefont {Bar~Elli}}, \bibinfo {author}
  {\bibfnamefont {Z.}~\bibnamefont {Fan}}, \bibinfo {author} {\bibfnamefont
  {A.~O.}\ \bibnamefont {Govorov}}, \ and\ \bibinfo {author} {\bibfnamefont
  {G.}~\bibnamefont {Markovich}},\ }\href {\doibase 10.1021/nl304638a}
  {\bibfield  {journal} {\bibinfo  {journal} {Nano Letters}\ }\textbf {\bibinfo
  {volume} {13}},\ \bibinfo {pages} {1203} (\bibinfo {year}
  {2013})}\BibitemShut {NoStop}%
\bibitem [{\citenamefont {Li}\ \emph {et~al.}(2022)\citenamefont {Li},
  \citenamefont {Gao}, \citenamefont {Zhang}, \citenamefont {Ji}, \citenamefont
  {Hu},\ and\ \citenamefont {Wu}}]{appMS2}%
  \BibitemOpen
  \bibfield  {author} {\bibinfo {author} {\bibfnamefont {H.}~\bibnamefont
  {Li}}, \bibinfo {author} {\bibfnamefont {X.}~\bibnamefont {Gao}}, \bibinfo
  {author} {\bibfnamefont {C.}~\bibnamefont {Zhang}}, \bibinfo {author}
  {\bibfnamefont {Y.}~\bibnamefont {Ji}}, \bibinfo {author} {\bibfnamefont
  {Z.}~\bibnamefont {Hu}}, \ and\ \bibinfo {author} {\bibfnamefont
  {X.}~\bibnamefont {Wu}},\ }\href {\doibase 10.3390/bios12110957} {\bibfield
  {journal} {\bibinfo  {journal} {Biosensors}\ }\textbf {\bibinfo {volume}
  {12}},\ \bibinfo {pages} {957} (\bibinfo {year} {2022})}\BibitemShut
  {NoStop}%
\bibitem [{\citenamefont {Goerlitzer}\ \emph {et~al.}(2023)\citenamefont
  {Goerlitzer}, \citenamefont {Zapata-Herrera}, \citenamefont {Ponomareva},
  \citenamefont {Feller}, \citenamefont {Garcia-Etxarri}, \citenamefont {Karg},
  \citenamefont {Aizpurua},\ and\ \citenamefont {Vogel}}]{appMS3}%
  \BibitemOpen
  \bibfield  {author} {\bibinfo {author} {\bibfnamefont {E.~S.~A.}\
  \bibnamefont {Goerlitzer}}, \bibinfo {author} {\bibfnamefont
  {M.}~\bibnamefont {Zapata-Herrera}}, \bibinfo {author} {\bibfnamefont
  {E.}~\bibnamefont {Ponomareva}}, \bibinfo {author} {\bibfnamefont
  {D.}~\bibnamefont {Feller}}, \bibinfo {author} {\bibfnamefont
  {A.}~\bibnamefont {Garcia-Etxarri}}, \bibinfo {author} {\bibfnamefont
  {M.}~\bibnamefont {Karg}}, \bibinfo {author} {\bibfnamefont {J.}~\bibnamefont
  {Aizpurua}}, \ and\ \bibinfo {author} {\bibfnamefont {N.}~\bibnamefont
  {Vogel}},\ }\href {\doibase 10.1021/acsphotonics.3c00174} {\bibfield
  {journal} {\bibinfo  {journal} {ACS Photonics}\ }\textbf {\bibinfo {volume}
  {10}},\ \bibinfo {pages} {1821} (\bibinfo {year} {2023})}\BibitemShut
  {NoStop}%
\bibitem [{\citenamefont {Yan}\ \emph {et~al.}(2014)\citenamefont {Yan},
  \citenamefont {Xu}, \citenamefont {Ma}, \citenamefont {Liu}, \citenamefont
  {Wang}, \citenamefont {Kuang},\ and\ \citenamefont {Xu}}]{appMS4}%
  \BibitemOpen
  \bibfield  {author} {\bibinfo {author} {\bibfnamefont {W.}~\bibnamefont
  {Yan}}, \bibinfo {author} {\bibfnamefont {L.}~\bibnamefont {Xu}}, \bibinfo
  {author} {\bibfnamefont {W.}~\bibnamefont {Ma}}, \bibinfo {author}
  {\bibfnamefont {L.}~\bibnamefont {Liu}}, \bibinfo {author} {\bibfnamefont
  {L.}~\bibnamefont {Wang}}, \bibinfo {author} {\bibfnamefont {H.}~\bibnamefont
  {Kuang}}, \ and\ \bibinfo {author} {\bibfnamefont {C.}~\bibnamefont {Xu}},\
  }\href {\doibase https://doi.org/10.1002/smll.201401641} {\bibfield
  {journal} {\bibinfo  {journal} {Small}\ }\textbf {\bibinfo {volume} {10}},\
  \bibinfo {pages} {4293} (\bibinfo {year} {2014})}\BibitemShut {NoStop}%
\bibitem [{\citenamefont {Mallat}\ \emph {et~al.}(2007)\citenamefont {Mallat},
  \citenamefont {Orglmeister},\ and\ \citenamefont {Baiker}}]{appCC1}%
  \BibitemOpen
  \bibfield  {author} {\bibinfo {author} {\bibfnamefont {T.}~\bibnamefont
  {Mallat}}, \bibinfo {author} {\bibfnamefont {E.}~\bibnamefont {Orglmeister}},
  \ and\ \bibinfo {author} {\bibfnamefont {A.}~\bibnamefont {Baiker}},\ }\href
  {\doibase 10.1021/cr0683663} {\bibfield  {journal} {\bibinfo  {journal}
  {Chemical Reviews}\ }\textbf {\bibinfo {volume} {107}},\ \bibinfo {pages}
  {4863} (\bibinfo {year} {2007})}\BibitemShut {NoStop}%
\bibitem [{\citenamefont {Li}\ \emph {et~al.}(2023)\citenamefont {Li},
  \citenamefont {Fan}, \citenamefont {Ye}, \citenamefont {Wu}, \citenamefont
  {Wang},\ and\ \citenamefont {Yin}}]{Science}%
  \BibitemOpen
  \bibfield  {author} {\bibinfo {author} {\bibfnamefont {Z.}~\bibnamefont
  {Li}}, \bibinfo {author} {\bibfnamefont {Q.}~\bibnamefont {Fan}}, \bibinfo
  {author} {\bibfnamefont {Z.}~\bibnamefont {Ye}}, \bibinfo {author}
  {\bibfnamefont {C.}~\bibnamefont {Wu}}, \bibinfo {author} {\bibfnamefont
  {Z.}~\bibnamefont {Wang}}, \ and\ \bibinfo {author} {\bibfnamefont
  {Y.}~\bibnamefont {Yin}},\ }\href {\doibase 10.1126/science.adg2657}
  {\bibfield  {journal} {\bibinfo  {journal} {Science}\ }\textbf {\bibinfo
  {volume} {380}},\ \bibinfo {pages} {1384} (\bibinfo {year}
  {2023})}\BibitemShut {NoStop}%
\bibitem [{\citenamefont {Singh}\ \emph {et~al.}(2014)\citenamefont {Singh},
  \citenamefont {Chan}, \citenamefont {Baskin}, \citenamefont {Gelman},
  \citenamefont {Repnin}, \citenamefont {Král},\ and\ \citenamefont
  {Klajn}}]{Science2014}%
  \BibitemOpen
  \bibfield  {author} {\bibinfo {author} {\bibfnamefont {G.}~\bibnamefont
  {Singh}}, \bibinfo {author} {\bibfnamefont {H.}~\bibnamefont {Chan}},
  \bibinfo {author} {\bibfnamefont {A.}~\bibnamefont {Baskin}}, \bibinfo
  {author} {\bibfnamefont {E.}~\bibnamefont {Gelman}}, \bibinfo {author}
  {\bibfnamefont {N.}~\bibnamefont {Repnin}}, \bibinfo {author} {\bibfnamefont
  {P.}~\bibnamefont {Král}}, \ and\ \bibinfo {author} {\bibfnamefont
  {R.}~\bibnamefont {Klajn}},\ }\href {\doibase 10.1126/science.1254132}
  {\bibfield  {journal} {\bibinfo  {journal} {Science}\ }\textbf {\bibinfo
  {volume} {345}},\ \bibinfo {pages} {1149} (\bibinfo {year}
  {2014})}\BibitemShut {NoStop}%
\bibitem [{\citenamefont {Jeong}\ \emph {et~al.}(2020)\citenamefont {Jeong},
  \citenamefont {Lee}, \citenamefont {Tran}, \citenamefont {Wang},
  \citenamefont {Lv}, \citenamefont {Park}, \citenamefont {Tang},\ and\
  \citenamefont {Lee}}]{ACSnano2020}%
  \BibitemOpen
  \bibfield  {author} {\bibinfo {author} {\bibfnamefont {K.-J.}\ \bibnamefont
  {Jeong}}, \bibinfo {author} {\bibfnamefont {D.~K.}\ \bibnamefont {Lee}},
  \bibinfo {author} {\bibfnamefont {V.~T.}\ \bibnamefont {Tran}}, \bibinfo
  {author} {\bibfnamefont {C.}~\bibnamefont {Wang}}, \bibinfo {author}
  {\bibfnamefont {J.}~\bibnamefont {Lv}}, \bibinfo {author} {\bibfnamefont
  {J.}~\bibnamefont {Park}}, \bibinfo {author} {\bibfnamefont {Z.}~\bibnamefont
  {Tang}}, \ and\ \bibinfo {author} {\bibfnamefont {J.}~\bibnamefont {Lee}},\
  }\href {\doibase 10.1021/acsnano.0c02026} {\bibfield  {journal} {\bibinfo
  {journal} {ACS Nano}\ }\textbf {\bibinfo {volume} {14}},\ \bibinfo {pages}
  {7152} (\bibinfo {year} {2020})}\BibitemShut {NoStop}%
\bibitem [{\citenamefont {Cheng}\ and\ \citenamefont {Sun}(2022)}]{couplex}%
  \BibitemOpen
  \bibfield  {author} {\bibinfo {author} {\bibfnamefont {Y.}~\bibnamefont
  {Cheng}}\ and\ \bibinfo {author} {\bibfnamefont {M.}~\bibnamefont {Sun}},\
  }\href {\doibase 10.1088/1367-2630/ac57e9} {\bibfield  {journal} {\bibinfo
  {journal} {New Journal of Physics}\ }\textbf {\bibinfo {volume} {24}},\
  \bibinfo {pages} {033026} (\bibinfo {year} {2022})}\BibitemShut {NoStop}%
\bibitem [{\citenamefont {Cheng}\ and\ \citenamefont {Sun}(2023)}]{coupley}%
  \BibitemOpen
  \bibfield  {author} {\bibinfo {author} {\bibfnamefont {Y.}~\bibnamefont
  {Cheng}}\ and\ \bibinfo {author} {\bibfnamefont {M.}~\bibnamefont {Sun}},\
  }\href {\doibase 10.1088/1367-2630/acc5a8} {\bibfield  {journal} {\bibinfo
  {journal} {New Journal of Physics}\ }\textbf {\bibinfo {volume} {25}},\
  \bibinfo {pages} {033028} (\bibinfo {year} {2023})}\BibitemShut {NoStop}%
\bibitem [{\citenamefont {Griffiths}(2013)}]{Griffiths}%
  \BibitemOpen
  \bibfield  {author} {\bibinfo {author} {\bibfnamefont {D.~J.}\ \bibnamefont
  {Griffiths}},\ }\href@noop {} {\emph {\bibinfo {title} {Introduction to
  Electrodynamics (4rd Edition)}}}\ (\bibinfo  {publisher} {Pearson},\ \bibinfo
  {address} {Cambridge, U.K.},\ \bibinfo {year} {2013})\BibitemShut {NoStop}%
\bibitem [{\citenamefont {Joe}\ \emph {et~al.}(2006)\citenamefont {Joe},
  \citenamefont {Satanin},\ and\ \citenamefont {Kim}}]{Fano}%
  \BibitemOpen
  \bibfield  {author} {\bibinfo {author} {\bibfnamefont {Y.~S.}\ \bibnamefont
  {Joe}}, \bibinfo {author} {\bibfnamefont {A.~M.}\ \bibnamefont {Satanin}}, \
  and\ \bibinfo {author} {\bibfnamefont {C.~S.}\ \bibnamefont {Kim}},\ }\href
  {\doibase 10.1088/0031-8949/74/2/020} {\bibfield  {journal} {\bibinfo
  {journal} {Physica Scripta}\ }\textbf {\bibinfo {volume} {74}},\ \bibinfo
  {pages} {259} (\bibinfo {year} {2006})}\BibitemShut {NoStop}%
\bibitem [{\citenamefont {Cheng}\ \emph {et~al.}(2018)\citenamefont {Cheng},
  \citenamefont {Zhang}, \citenamefont {Zhao}, \citenamefont {Wen},
  \citenamefont {Hu}, \citenamefont {Gong},\ and\ \citenamefont
  {Lu}}]{Cheng2018}%
  \BibitemOpen
  \bibfield  {author} {\bibinfo {author} {\bibfnamefont {Y.}~\bibnamefont
  {Cheng}}, \bibinfo {author} {\bibfnamefont {W.}~\bibnamefont {Zhang}},
  \bibinfo {author} {\bibfnamefont {J.}~\bibnamefont {Zhao}}, \bibinfo {author}
  {\bibfnamefont {T.}~\bibnamefont {Wen}}, \bibinfo {author} {\bibfnamefont
  {A.}~\bibnamefont {Hu}}, \bibinfo {author} {\bibfnamefont {Q.}~\bibnamefont
  {Gong}}, \ and\ \bibinfo {author} {\bibfnamefont {G.}~\bibnamefont {Lu}},\
  }\href {\doibase 10.1088/1361-6528/aac44f} {\bibfield  {journal} {\bibinfo
  {journal} {Nanotechnology}\ }\textbf {\bibinfo {volume} {29}},\ \bibinfo
  {pages} {315201} (\bibinfo {year} {2018})}\BibitemShut {NoStop}%
\end{thebibliography}%

\end{document}